\documentclass[12 pt, amssymb,prb,showpacs]{revtex4}
\usepackage{epsfig}
\usepackage{dcolumn}
\usepackage{amsmath}
\usepackage[a4paper,dvips]{geometry}
\begin{document}
\title{\bf Plasmon phenomena as origin of DC-current induced
resistivity oscillations in two-dimensional electron systems}
\author{Jes\'us I\~narrea}
 \affiliation {Escuela Polit\'ecnica
Superior,Universidad Carlos III,Leganes,Madrid,Spain and  \\
Instituto de Ciencia de Materiales, CSIC,
Cantoblanco,Madrid,28049,Spain.}
\date{\today}
\begin{abstract}
We analyze theoretically the oscillations that the
magnetoresistivity of two-dimensional electron systems present when
a high intensity direct current is applied. In the model presented
here we suggest that a plasma wave is excited in the system
producing an oscillating motion of the whole two-dimensional
electron gas at the plasma frequency. This scenario affects
dramatically the way that electrons interact with scatterers giving
rise to oscillations in the longitudinal resistivity. With this
theoretical model experimental results are well reproduced and
explained.

\end{abstract}
\maketitle
\newpage
Magnetotransport properties of highly mobile two-dimensional
electron systems (2DES) is a subject of increasing interest.
The future nanoelectronics will integrate 2DES as key parts of
different devices as transistors, diodes or capacitors. In
particular the phenomena of microwave-induced resistivity
oscillations (MIRO) and zero resistance states
(ZRS)\cite{mani,zudov2,studenikin,islamov,dietel, kunold,auer,mani2}
have attracted considerable attention both from theoretical and
experimental sides. However the mechanism responsible is still under
debate.

Another striking non-linear effect has been observed when a 2DES is
subjected to a high intensity DC-current\cite{yang,bykov,zhang}.
These experiments report oscillations in the longitudinal
magnetoresistivity ($\rho_{xx}$) that are periodic in inverse
magnetic field and tend to be qualitatively similar to the ones
obtained in the experiments of microwave (MW) excited
magnetoresitivity in Hall bars\cite{mani,zudov2,studenikin,islamov}.
Another remarkable experimental outcome is that under DC-current
excitation, the $\rho_{xx}$ response shifts towards higher magnetic
fields ($B$) with increasing DC-current intensity. Also the electron
sheet density dependence shows that the maxima positions in the
$\rho_{xx}$ oscillations are scaled as $1/\sqrt{n_{e}}$ where
$n_{e}$ is the electron sheet density. The first explanation to
these effects\cite{yang,bykov,zhang} considers that oscillations are
related to elastic scattering mediated by a short range disorder
between Hall field-tilted Landau levels. Accordingly an electron may
transfer momentum in the x-coordinate that is equivalent to a
hopping in the y direction. $\rho_{xx}$ reflects this hopping
through the conductivity $\sigma_{yy}$. Other theoretical proposals
have been reported recently\cite{lei2}.

In this letter we present an alternative approach to explain these
effects. In our model the DC-electric field ($E_{DC}$), which
produces the DC-current ($I_{DC}$), displaces the orbit center of
the harmonic quantum oscillators that are the electrons in the
presence of a uniform and perpendicular $B$. This displacement with
respect to the positive lattice ions disturbs  the 2D electron gas
leading to a collective excitation of the system. In other words, a
plasma wave originates from a spatial shift of the 2DES as the
charge seeks to move to restore equilibrium. The system responds
with an oscillatory motion where the frequency is denoted as $w_{p}$
or plasma frequency. Evidently its effect is only measurable at
sufficient high DC-current intensities. This scenario affects
dramatically the way that electrons interact with scatterers and is
reflected in the $\rho_{xx}$ response. We explain and reproduce the
$\rho_{xx}$ oscillations periodicity in $B^{-1}$, the shift to
larger $B$ with increasing current intensity and the dependence on
$n_{e}$.

The electrons of a 2DES subjected to a perpendicular $B$, behave as
harmonic quantum oscillators. Then if we apply a constant electric
field $E_{DC}$ in the current direction (x-direction), we obtain a
displacement $x$ of the center of all oscillators in the same
direction: the entire 2D electron gas moves. The maximum value of
$x$ is given by\cite{ina,ina2}
$x_{max}=\frac{eE_{DC}}{m^{*}w_{c}^{2}}$ where $e$ is the electron
charge and $w_{c}$ the cyclotron frequency. The spatial shift of the
electron gas as a whole with respect to the fixed positive
background (lattice ions) give rise to two lines of opposite charge
at either end of the sample creating an electric field $E_{p}$(see
Fig. 1). We obtain this field using basic Electrostatics\cite{wag}:
$\overrightarrow{E}(p)=\frac{n_{e} e x}{2\pi \epsilon
L_{ef}}\overrightarrow{i}$
being $L_{ef}$ the average effective distance between the two lines
of opposite charge that we approximate by $L_{ef}=L-x_{max}$,
$\epsilon$ is the dielectric constant and $L$ the sample length.
This field tends to restore the system to its equilibrium position
producing in the system a collective excitation or plasma wave.
Eventually the 2D electron gas obey the equation of motion of a
harmonic oscillator\cite{ash} that leads to oscillations at the
plasma frequency $w_{p}$:
\begin{equation}
w_p=\sqrt{\frac{n_{e} e^{2}}{2\pi m^{*}\epsilon L_{ef}}}
\end{equation}
On the other hand the plasma oscillation can be damped. Physically
the coupling of the plasma to the lattice through phonon scattering
results in a net decrease of the energy of the oscillating electron
gas. The result is a progressive decay of the plasma oscillation
being reflected in a continuous decrease of its amplitude.
Eventually the plasma oscillations can be totally damped. However
the external field  $E_{DC}$ is still acting on the 2DES exciting
the plasma wave and avoiding the oscillatory motion to stop. Thus
$E_{DC}$ plays the role of and exciting force on the oscillating
plasma. If we name this force as $F_{DC}$, the plasma equation of
motion can be written as:
\begin{equation}
Nm^{*}\frac{d^{2}x}{dt}+Nm^{*}\gamma \frac{dx}{dt}+Nm^{*} w_{p}^{2}x
=F_{DC}
\end{equation}
where $N$ is the number of electrons and $\gamma$ is a
phenomenologically introduced damping factor which accounts for the
electronic interactions with the lattice yielding acoustic phonons.
To solve this differential equation we propose as a possible
solution a harmonic function like $x_{p}=A_{DC}\cos w_{p}t$. The
expression of $F_{DC}$ and the amplitude $A_{DC}$ are to be
determined. Thus we obtain:
\begin{equation}
 x_{p}=\frac{eE_{DC}}{m^{*}\gamma w_{p}}\cos w_{p}t
\end{equation}
if  $F_{DC}=NeE_{DC}\cos wt$  and $ A_{DC}=
\frac{eE_{DC}}{m^{*}\gamma w_{p}}$.
Therefore we have a 2DES that oscillates as a whole with $w_{p}$,
displacing the electrons orbit center with $x_{p}(t)$. The
oscillations are excited by $F_{DC}$ and damped by coupling with the
lattice.

Now we introduce the scattering suffered by electrons due to charged
impurities randomly distributed in the sample. Without high
intensity DC-current excitation, an electron in an initial state
corresponding to an orbit center position $X_{n}^{0}$, scatters and
jumps to a final state with orbit center $X_{m}^{0}$, changing its
average coordinate in the static electric field direction in $\Delta
X^{0}=X_{m}^{0}-X_{n}^{0}$. This is the relevant direction
(x-direction) in this problem due to the presence of the DC electric
field. It means that all the scattering jumps will be averaged to
zero except the ones in the x-direction. Under a high intensity
DC-current the plasma wave is excited and the electronic orbit
center coordinates
change being given, according to our model\cite{ina,ina2}, by: \\
$X^{p}=X^{0}+x_{p}=X^{0}+A_{DC}\cos w_{p}t$. Thus, due to the
collective oscillation all the electronic orbit centers in the
sample oscillate back and forth in the x direction through $x_{p}$.
When an electron suffers a scattering process with a probability
rate given by $W_{m,n}$, it takes a time $\tau = \frac{1}{W_{m,n}}$
for that electron to jump from an orbit to another. This probability
rate is calculated according to the model described in
ref\cite{ina,ina2}. If we consider that the electron jumps from the
initial state oscillation $middle$ position, and it takes a time
$\tau$ to get to the final one, then we can write for the average
coordinate change in the x direction:
$\Delta X^{p}=\Delta X^{0}+ A_{DC}\cos w_{p}\tau$
Finally the longitudinal conductivity $\sigma_{xx}$ can be
calculated: $\sigma_{xx}\propto \int dE \frac{\Delta
X^{p}}{\tau}(f_{i}-f_{f})$,  being $f_{i}$ and $f_{f}$ the
corresponding electron distribution functions for the initial and
final states respectively and $E$ energy. To obtain $\rho_{xx}$ we
use the well-known tensor relation
$\rho_{xx}=\frac{\sigma_{xx}}{\sigma_{xx}^{2}+\sigma_{xy}^{2}}$,
where $\sigma_{xy}\simeq\frac{n_{e}e}{B}$.

In Fig. 2a, we present calculated $\rho_{xx}$ versus $B$ at a direct
current $I_{DC}=300\mu A$. We observe clear oscillations in
$\rho_{xx}$ showing four peaks. The oscillations can be explained in
similar terms as the ones obtained with MW excitation\cite{ina}(see
Fig. 3). When no plasma wave is excited, electrons jump between
fixed orbits and on average an electrons advances a distance $\Delta
X^{0}$ (Fig. 3.a). When the DC-current excitation is on, orbits are
not fixed and instead move back and forth through $x_{p}$ with the
frequency $w_{p}$. When the orbits, due to plasma oscillation, are
moving backwards during the scattering jump the electrons advance an
average larger distance than the no plasma wave case: $\Delta
X^{p}>\Delta X^{0}$. This corresponds to an increasing conductivity
(Fig. 3b). When the entire electron gas is moving forward, during
the jump the electron advances on average a shorter distance:
$\Delta X^{p}<\Delta X^{0}$ (Fig. 3c). This corresponds to a
decrease in the conductivity with respect to the case without
DC-current excitation.

In Fig. 2b we present calculated $\rho_{xx}$ versus $B^{-1}$. We
observe that $\rho_{xx}$ is periodic in $B^{-1}$ in agreement with
experiments\cite{yang,bykov,zhang} with an spatial period of
$\delta$, (see figure). This is clearly shown in the inset where we
present the inverse of peaks position, $B^{-1}_{n}$ versus the order
of the peaks $n$. According to our model, $\rho_{xx}\propto cos
w_{p}\tau = cos \frac{w_{p}}{K B}$\cite{ina}, $K$ being a constant,
i.e., $\rho_{xx}$ is periodic with $B^{-1}$. Thus, in the peaks the
next condition is fulfilled: $B^{-1}=\frac{K}{w_{p}}2\pi n$.
Therefore we obtain the equation of a straight line that crosses the
origin as in experiments\cite{yang,bykov}.

In Fig. 4, we present calculated $\rho_{xx}$ versus $B$ for
different values of $I_{DC}$. We observe that the peaks shift
towards higher $B$ with increasing $I_{DC}$. An increasing $I_{DC}$
corresponds to a larger $E_{DC}$ and $x_{max}$ affecting eventually
$w_{p}$ (see eq. 2) that becomes larger too. Considering the
functional dependence of $\rho_{xx}$ through $\cos w_{p} \tau$, a
larger $w_{p}$ give rise to more peaks and a shift of $\rho_{xx}$
response to larger $B$. A similar behavior is obtained in MW-excited
$\rho_{xx}$ response in Hall bars\cite{mani}. In Fig. 5, we present
the calculated peaks position $B_{n}$ versus $n_{e}^{1/2}$. Two
straight lines crossing the origin are obtained for peak order 1 and
2. The inset shows $\rho_{xx}$ versus $B$ for different $n_{e}$.
Again following our model, $\rho_{xx} \propto cos
\frac{w_{p}}{W_{m,n}} = cos C\frac{n_{e}^{1/2}}{n_{e}B}=cos
\frac{C}{n_{e}^{1/2}B}$\cite{ina}, $C$ being a constant. Peaks
maxima fulfill  $2\pi n=\frac{C}{n_{e}^{1/2}B}\Rightarrow
B=\frac{C}{2\pi n}\frac{1}{n_{e}^{1/2}}$. As in the experimental
outcome\cite{yang}, we obtain the equation of a straight line
crossing the origin.

This work has been supported by the MCYT (Spain) under grant
MAT2005-06444, by the Ram\'on y Cajal program and  by the EU Human
Potential Programme: HPRN-CT-2000-00144.

\newpage
\clearpage

Fig.1 caption: Schematic diagram showing electronic transport
through a Hall Bar. A high intensity DC-current produces a
displacement $x_{max}$ of electrons orbit center with respect to the
fixed positive background of lattice ions. The resulting linear
distributions of opposite charge in opposite sides of the 2D sample
give rise to an electric field $E_{P}$.
\newline

Fig.2 caption: (a) Calculated $\rho_{xx}$ versus $B$ at a direct
current $I_{DC}=300\mu A$. We observe clear oscillations in
$\rho_{xx}$ response. Peaks order from 1 to 4 is shown. (b)
Calculated $\rho_{xx}$ versus inverse magnetic field ($B^{-1}$). The
inset shows that $\rho_{xx}$ is periodic in $B^{-1}$, being $\delta$
the spatial period. T=1K.
\newline

Fig.3 caption: Schematic diagrams of electronic transport through
the 2D sample without (a), (fixed orbits) and with plasmonic
excitation (b) and (c), (oscillating orbits).
\newline

Fig. 4 caption: Calculated $\rho_{xx}$ versus $B$ for different
values of $I_{DC}$ ($50\mu A \rightarrow 400\mu A$). We observe, as
in experiments, that the peaks shift towards higher $B$ with
increasing $I_{DC}$. The $\rho_{xx}$  shift and the increasing
number of peaks with  $I_{DC}$ suggest and increasing $w_{p}$. T=1K.
\newline

Fig. 5 caption: Calculated peak position $B_{n}$ versus
$n_{e}^{1/2}$. $n_{e}$ is the electron sheet density. Two straight
lines crossing the origin are obtained for peak order 1 and 2. The
inset shows $\rho_{xx}$ versus $B$ for different $n_{e}$. T=1K.


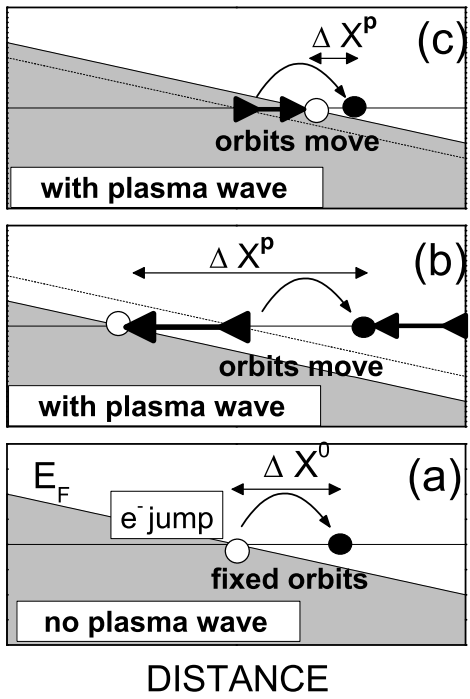
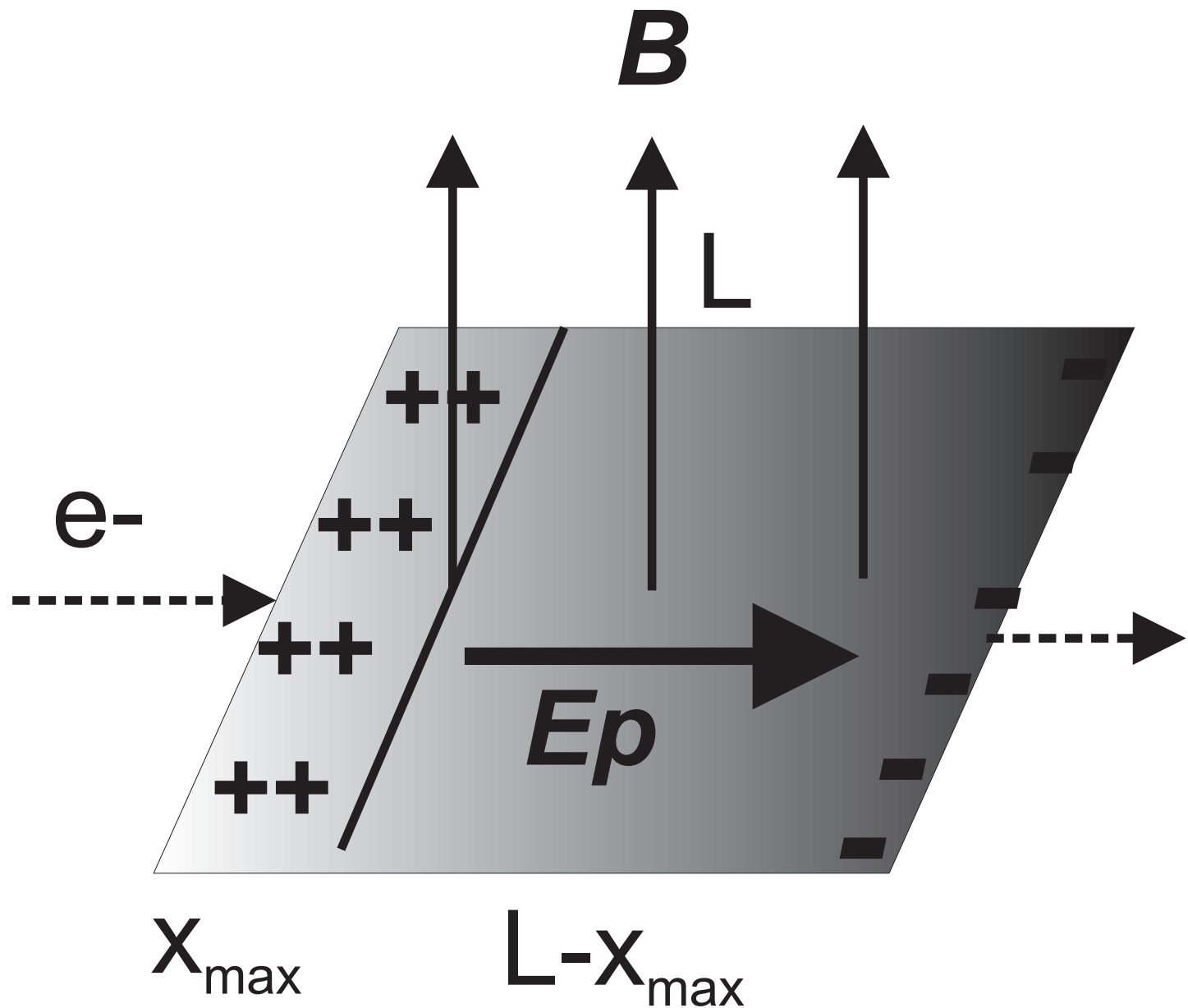
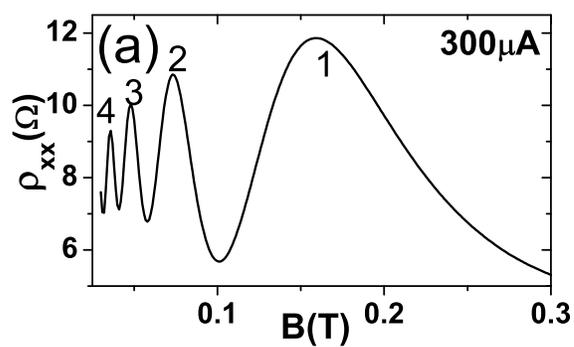
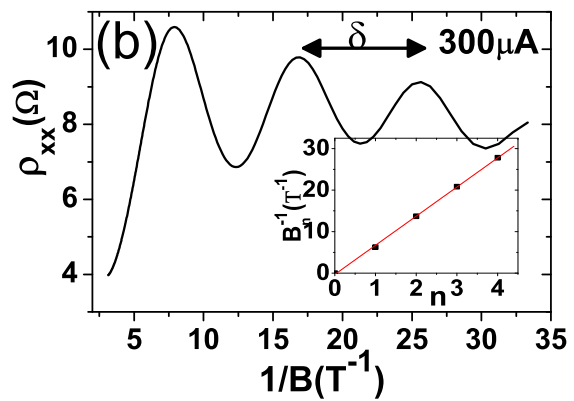
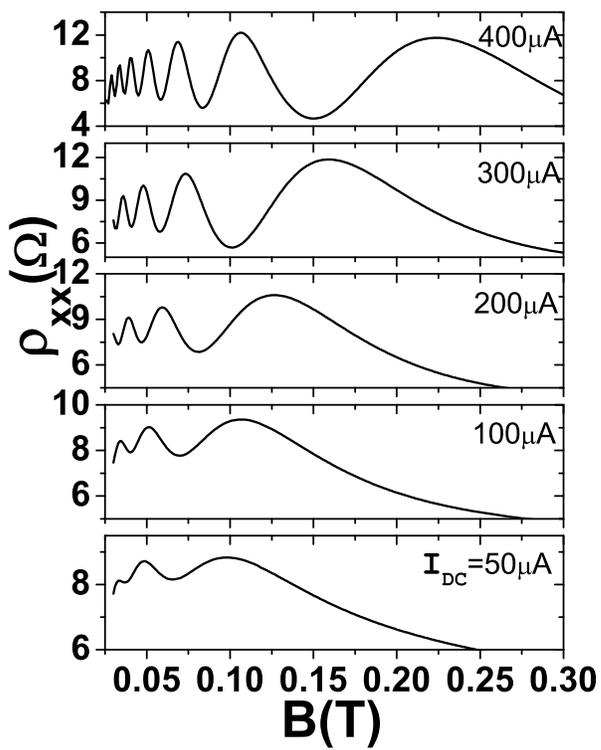
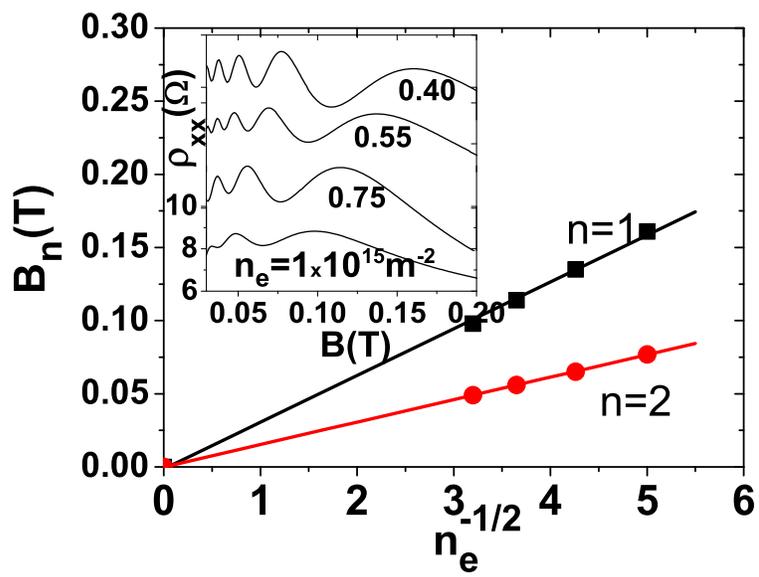
\end{document}